# Towards a collaborative digital platform for railway infrastructure projects

Pierre JEHEL[1], Pierre-Étienne GAUTIER[1], Judicaël DEHOTIN[2], Flavien VIGUIER[2]

[1] Université Paris-Saclay, CentraleSupélec, ENS Paris-Saclay, CNRS, LMPS - Laboratoire de Mécanique Paris-Saclay, 91190, Gif-sur-Yvette, France
[2] SNCF Réseau, Industrial Engineering direction, 94574, La-Plaine-Saint-Denis, France
Corresponding Author: Pierre Jehel (pierre.jehel@centralesupelec.fr)

**Abstract**

The management of railway infrastructure projects can be supported by collaborative digital platforms. A survey was carried out to identify the needs and expectations of the various stakeholders involved in the design and construction of railway infrastructure projects regarding collaborative platforms. These needs and expectations can then be translated into functional specifications to be included in the digital platforms. A total of 21 interviews were conducted between October and December 2022, during which 35 individuals were interviewed. Key roles were represented across the different project phases: engineers from design and construction firms, project managers, infrastructure managers. And various engineering fields were represented: civil, electrical, telecommunications, tracks, systems. These interviews were carried out by CentraleSupélec | Université Paris-Saclay and by SNCF Réseau using a structured protocol designed to collect the specific needs of the interviewees for collaboration, as well as the guiding principles that shape both individual work practices and collaboration between professions. The resulting material was analyzed and then synthesized into a conceptual model of a collaborative digital platform for supporting the design and construction phases in a railway infrastructure project. Also, from these interviews emerged five core functionalities that the platform must offer: Providing access to existing infrastructure data; Accelerating repetitive tasks; Verifying essential project requirements; Supporting decision-making; Facilitating coordination among stakeholders.

Keywords: cross-sectorial collaboration, design and construction, field interviews, digitization, digital twin

## 1. Introduction

A great many things depend on the speed of a train. Travel times offered to passengers of course. But also, the curves and slopes of the tracks, the electric traction system, the design of the piles in a bridge, the signaling and telecommunications systems, and so on. Behind each of these specific concepts, there are engineers who were trained for different specific professions (civil, electrical, computer, systems… engineering). Each profession contributes to and is essential to the realization of a railway infrastructure project in the domain of their specific expertise. And each profession also has interactions with professions from outside of their domain. For instance, if a track engineer assigns a slope to a track portion, their design has to be shared with electrical engineers because it implies a compliant electric traction design.

Each profession has developed specific practices and habits that make them highly efficient within their domain. Each profession has also built the capacity to operate beyond their own domain to adapt their design to constraints coming from other domains. Going back to our example above, it is likely that an optimal track profile from the track engineers point of view is not an optimal one from the viewpoint of the electrical traction engineers. And this would also be likely with the signaling system engineers, the telecommunication system engineers, the civil engineers, and so on, thus making the overall process of designing and building a railway infrastructure potentially complex.

Managing a railway infrastructure project therefore falls into the category of controlling the evolution of a complex system. Consequently, project managers and infrastructure owners have been investing in research and development to manage the complexity across the sector and all along the design and construction process until delivering projects that meet the levels of performance required. An example of such investments in R&D is the Minerve project in France where infrastructure managers have built a research consortium together with construction and design companies, and with CentraleSupélec | Université Paris-Saclay as academic partner. The work presented hereafter has been carried out within the framework of this Minerve project.

Minerve aims at fostering and developing digital tools for managing complexity in railway infrastructure projects. Formulating specifications for a collaborative digital platform is one of the objectives followed by Minerve. In this paper, we report on a series of interviews that were conducted to learn about the services that future users would expect to be provided by a collaborative digital platform. The main conclusion is that the platform should integrate the essential constraints that must be satisfied throughout the infrastructure design and construction project and that it should support mediations between the people from various engineering domains that contribute to the project.

How the interviews were conducted is presented in the next section. Then, section 3 introduces the methodology that was used to analyze the content of the interviews. In section 4, a synthesis of the interviews is proposed showing a conceptual model of collaborative digital platform and five core functionalities that this platform should offer to its users. Some concluding remarks close the paper.

## 2. Conducting the interviews

21 interviews were conducted between October and December 2022 gathering 35 interviewees representing various expert domains (electric traction, signaling, telecommunications, bridge and tunnel engineering, railway tracks engineering, systems engineering) from various stakeholders (infrastructure managers, design companies, construction companies). Each interview lasted approximately 50 minutes. All discussions were recorded and stored. Consent to record, store, and make the interviews available to the Minerve project was obtained from each interviewee.

A questionnaire was prepared prior to the interview. It is organized into four sections designed to explore various topics and progressively broaden the scope, from the specific roles of the interviewees to more general considerations regarding the integration of these roles within a collaborative digital approach: (i) Professional role identification; (ii) Collaborative methodologies already adopted; (iii) Experience with common data environments; (iv) General ideas about collaborating in a railway infrastructure project. The questionnaire served as a conversational guide. In some interviews, all questions were addressed, in others only a few of them. The initial question of each section of the questionnaire was always asked to initiate the discussion. Each of the four sections was then systematically explored, though not all individual questions were necessarily covered, as priority was given to the free expression of the participants.

This empirical approach aimed to create space for innovation within the conversation, allowing participants to speak not only about the specific tasks they perform in their daily work, but also about the imagined possibilities that emerge when thinking of digitizing those activities.

## 3. Analysis of the interviews

### 3.1 Methodology

In (Latour, 2012), it is demonstrated that the continuity of systems is ensured by a series of discontinuities (for example, the replacement of worn mechanical pieces, while generating discontinuities in the process, ensures the continuity of a production line). Accordingly, we assume that ensuring digital continuity in a process means integrating and carrying through a series of discontinuities during this process. Our interview analysis is therefore based on detecting discontinuities in the interviewees' discourse. These breaks in the continuity of speech – these hiatuses as they are called in (Latour, 2012) – reveal what people and the professions they represent truly care about, their needs, and what must be preserved in a collaborative digital platform. These hiatuses also reveal the guiding forces behind professional practices, that are the elements people rely on and navigate through to overcome or blur discontinuities.

### 3.2 What experts expect to find in a collaborative digital platform

Below is a list of some of the discontinuities detected on several occurrences during the interviews. These often appear in the form of a "but" linking two sentences.

"The issue in collaborative work is more human than technological. But a system without humans could not properly function." This means that a collaborative digital platform would not guarantee people would collaborate. Consequently, the platform should be designed to support collaboration providing services needed by one profession to work in interaction with the others.

"We exchange data, but the other professions use another format, so we do not understand each other." Data is often exchanged in incompatible formats, making comparison and integration difficult. The platform should rely on formats every collaborator could use.

"A high level of detail in object geometry and positioning helps anticipate compatibility issues between expert professions. But it also creates unnecessary problems because we share highly uncertain data that we know will change during the project." One interviewee mentioned that before digital tools, hand drawings on tracing paper overlays were used, that these drawings only included essential details and that yet it worked. The collaborative platform should manage the level of reliability of the data it contains.

"We need exhaustive and reliable data. But we don't have a sufficiently developed common data environment (CDE) to manage it." Common data environment work more like folders where files are stored than like data sources.

"We have an electronic documentation process, but there are several document management systems, and we have to search through them to find the right information." This raises the issue of who is in charge of the platform, that is who chose the platform and make sure every stakeholder can interact with it, where the common data and algorithms are stored, and so on.

### 3.3 Channels for information flow

We identified the guiding forces that allow professionals to move past the discontinuities they experience and go on with the course of their work. These guiding forces channel and conduct information flowing through the project. Like pipes that can only carry what their diameter allows, these forces filter what can flow through. It is important a digital platform integrates this information channels as they are key to collaboration. Below is an excerpt of the guiding forces identified in the analysis of the interviews.

**Habits.** Habits help experts being precise, reliable, and efficient. They also let them move past a discontinuity in their processes: the experts that experience it know what to do because they are used to it.

**The law / contracts.** Contracts define responsibilities that shape processes and information flows within the project. A contradiction stops being one if it does not break contractual rules.

**Networks**. "I find the data and information I need by reaching out to colleagues out of my domain." Relying on networks, inside and outside one's organization, is integral to professional life.

### 4. Synthesis of the interviews

### 4.1 Notations

We call the collaborative digital platform PARIS (Platform for Augmented Railway Infrastructure System) and we refer to {PARIS} as a virtual representation of the real-world railway infrastructure system {INFRA}. {INFRA} is described by a digital state $X$ that consists of a set of digital features sufficient to characterize the system.

### 4.2 Conceptual model for a collaborative digital platform

A conceptual model of PARIS is proposed in Figures 1 and 2. We started from the state-of-the-art conceptual model of a digital twin presented in (Thelen *et al.*, 2022) and enriched it with a decision-making process that can notably integrate human components, thus affirming that the collaborative digital platform PARIS acts as a decision support tool to command actions on the real system {INFRA}.

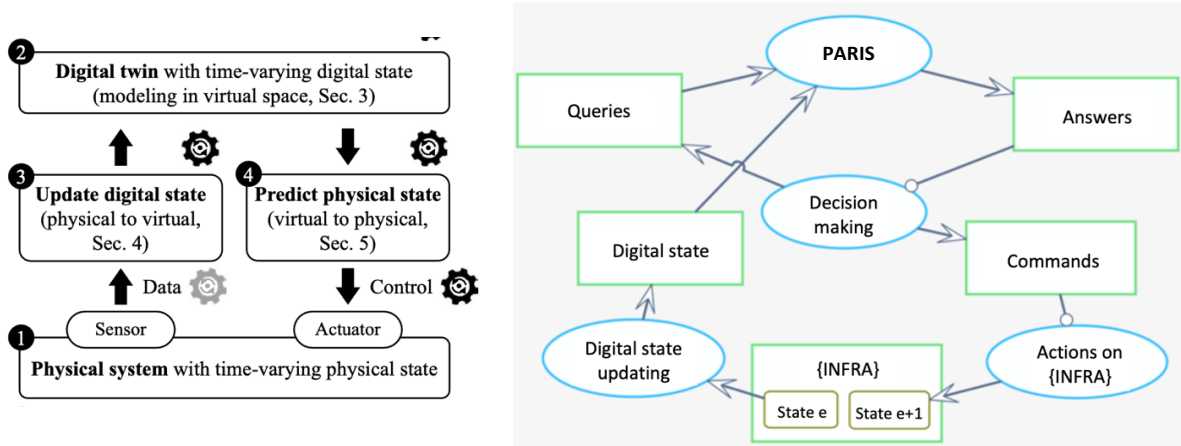

**Figure 1:** [Left] Conceptual model of a digital twin (Thelen *et al.*, 2022); [Right] Proposed conceptual model to connect {INFRA} and PARIS. The representation follows the Object-Process Methodology (OPM) (screenshot from the OPCloud trial version). In the OPM formalism (Dori, 2016), ovals represent processes, and rectangles represent objects involved in the processes. The processes depicted here require resources (lines connected to the processes by circles) and generate an object (unidirectional arrows).

Figure 2 presents a close-up view of the collaborative platform PARIS. Human collaboration is represented at different levels: within the mediation process between the experts and within the various processes internal to the professions. These processes benefit from the assistance provided by {PARIS} that acts as a digital twin of the real system {INFRA}.

{INFRA} evolves over time. {PARIS} receives a description of {INFRA} that is in a state *e* represented by a set of features gathered in $X_e$. To answer queries about {INFRA}, a design engineering work is organized with usually more than one profession involved. To move from a state $X_e$ to the next state $X_{e+1}$, the different professions *p* first identify, in a mediation process, the features of $X_e$ which they are responsible to make evolve ($X_{1,1}^{(1)}$ for profession $p = 1$, $X_{1,1}^{(2)}$ for $p = 2$ …). Then, each profession works with their features and estimate how they should be modified to answer the query they received about {INFRA}. At the next mediation, the experts gather and present updated features forming a new state $X_{1,1}$. This new state is a priori not final: it may contain inconsistencies because different experts may have worked on common features and made them evolve differently. The mediation results in an agreement on how to update the features. Each profession then works again with the new features ($X_{1,2}^{(1)}$, $X_{1,1}^{(2)}$,…). And so on until obtaining a state $X_{1,N} = X_2$ that satisfies all the experts and that answers the query is reached.

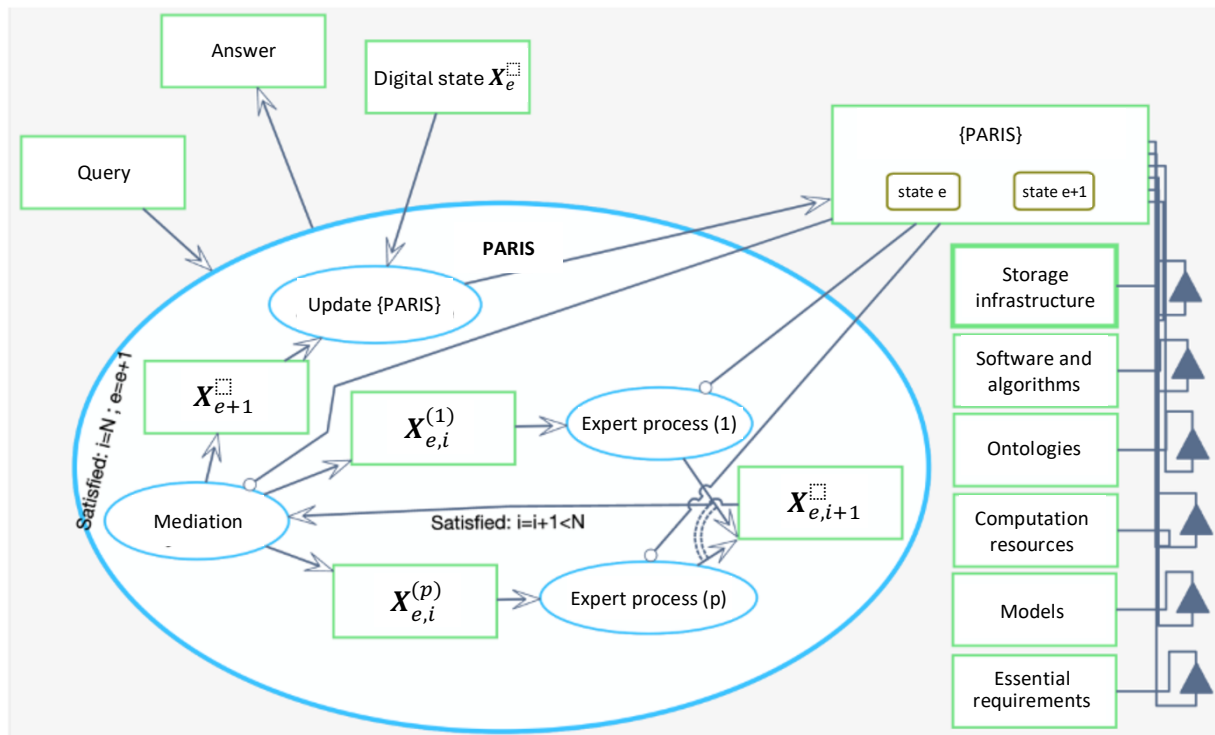

**Figure 2:** Close-up view of the collaborative platform. A query is received about the real system {INFRA}. {INFRA} is described by a digital state $X_e$. Assisted by {PARIS}, various professions $p$ (experts) collaborate and make the digital state evolve until converging to an answer to the query.

The conceptual model of the collaborative platform explicitly introduces a human organization and gives it a central place. The human organization is assisted by {PARIS}, the virtual representation of the physical system {INFRA}. In {PARIS}, there are common frameworks and data for collaboration. Data in {PARIS} are updated each time all the stakeholders have agreed on an evolution of the state of the system. {PARIS} is therefore updated both with data about the human processes and with data recorded on the physical system {INFRA}.

The question of which stakeholder in the project is responsible for the platform remains open (where are common data, algorithms, models… stored? Who administrates the platform? …).

**4.3 Five functionalities the collaborative digital platform must offer**

From the analysis of the interviews, we concluded that a collaborative digital platform should be:

**Providing access to the existing data**. Any stakeholder joining the project should be able to find the information they need from the other stakeholders and to share the information the others expect from them.

**Accelerating the execution of repetitive tasks**. A repetitive task is, for the person performing it, a task they would prefer to do only once or not at all if it has already been done elsewhere.

**Verifying essential requirements.** An essential requirement is a sine qua non condition for the success of the project, whether operational, contractual, regulatory, legislative,… An essential requirement may not fully satisfy one of the project stakeholders; it then represents the extent to which they are willing to go to guarantee the project's success without jeopardizing their own activities.

**Managing uncertainties.** Uncertainty can arise from a lack of information, errors, inability to precisely predict the consequences of an action, hazards, natural variability… Managing uncertainties allows each profession to control the tolerance range within which they can operate to satisfy their own requirements while offering room for maneuver to other professions with which they are interdependent.

**Fostering mediations.** A mediation occurs when two people working in different domains feel the need to exchange information in a mutually enriching way.

## 5. Conclusions

The growing complexity of railway systems demands integrated and collaborative engineering practices to ensure infrastructure and rolling stock robustness, reliability, and safety. The integration of advanced digital technologies is redefining interactions between technical stakeholders and other project participants. Several organizations report a reduction in technical conflicts through the adoption of such approaches. However, tool heterogeneity and the management of information flows remain major challenges.

Besides, the convergence of model-based system engineering (MBSE), building information modeling (BIM), cloud computing, and generative AI suggests the emergence of digital systems capable of managing complex projects by efficiently orchestrating both human and algorithmic contributions. However, successful adoption requires a simultaneous transformation of business processes, technical skills, and organizational governance.

This paper presents a conceptual model and key functionalities to develop collaborative digital platforms that could support managing the complexity inherent to designing and building railway infrastructures and that could also directly support the engineering practice.

## Acknowledgment

The results presented in this paper come from the analyses and synthesis of interviews that were conducted in the French railway sector within Minerve, a research project funded by the French Government in the France 2030 framework. The authors thank the 35 interviewees and their companies.